\documentclass{article}
\usepackage[T1]{fontenc}
\usepackage[utf8]{inputenc}
\usepackage[lbd]{ismir} 
\usepackage{amsmath,cite,url}
\usepackage{graphicx}
\usepackage{color}
\usepackage{booktabs}
\usepackage{placeins}
\usepackage{amssymb}
\usepackage{pifont}

\usepackage{algorithm} 
\usepackage[noend]{algpseudocode} 
\usepackage{gensymb} 
\usepackage{siunitx} 
\usepackage{multirow} 
\usepackage{siunitx} 
\usepackage{kotex}
\usepackage{microtype} 
\usepackage{paralist} 
\usepackage{float} 

\title{Two Web Toolkits for Multimodal Piano Performance Dataset Acquisition and Fingering Annotation} 

\multauthor
  {Junhyung Park$^\flat$ \hspace{1cm} Yonghyun Kim$^\natural$ \hspace{1cm} Joonhyung Bae$^\sharp$ \hspace{1cm} Kirak Kim$^\sharp$}
  {{\bfseries Taegyun Kwon$^\sharp$ \hspace{1cm} Alexander Lerch$^\natural$ \hspace{1cm} Juhan Nam$^\sharp$}\\
  $^\flat$ Department of Mathematical Sciences, KAIST, South Korea\\
  $^\natural$ Music Informatics Group, Georgia Institute of Technology, USA\\
  $^\sharp$ Graduate School of Culture Technology, KAIST, South Korea\\
  {\tt\small \{tonyishappy, jh.bae, kirak, ilcobo2, juhan.nam\}@kaist.ac.kr,}\\
  {\tt\small \{yonghyun.kim, alexander.lerch\}@gatech.edu}
  }
  
\def\authorname{J. Park, Y. Kim, J. Bae, T. Kwon, K. Kim, A. Lerch and J. Nam}

\begin{document}

\maketitle 

\begin{abstract}



Piano performance is a multimodal activity that intrinsically combines physical actions with the acoustic rendition. Despite growing research interest in analyzing the multimodal nature of piano performance, the laborious process of acquiring large-scale multimodal data remains a significant bottleneck, hindering progress in this field. To overcome this barrier, we present an integrated web toolkit comprising two Graphical User Interfaces (GUIs):
\begin{inparaenum}[(i)]
\item \textit{PiaRec}, which supports the synchronized acquisition of audio, video, MIDI, and performance metadata, and
\item \textit{ASDF}, which enables the efficient annotation of performer fingering from the visual data. 
\end{inparaenum}
Collectively, these tools streamline the acquisition of multimodal piano performance datasets.

\end{abstract}

\section{Introduction}\label{sec:introduction}


The computational study of piano performance as a multimodal activity offers deep insights into musical artistry and technique \cite{jensen2012multimodal, riley2005use, parmar2021piano}. Thus, multimodal piano datasets combining audio, video, MIDI and fingering annotations play a crucial role for understanding piano performance. However, existing acquisition methods often require manual synchronization across multiple software tools and fingering annotation by experts, limiting dataset scale and accessibility.

This challenge is especially prominent for fingering data. While fundamental to performance technique, the high degree of subjectivity in fingering makes it difficult to collect and analyze systematically \cite{swinkin2007keyboard}. To address these dataset acquisition and fingering annotation challenges, we introduce an integrated web toolkit, which consists of \textit{PiaRec} and \textit{ASDF} (semi-Automated System for Detecting Fingering).\footnote{\url{https://github.com/yonghyunk1m/PianoVAM-Code}} PiaRec automates the synchronized recording of multimodal data, while ASDF provides an efficient human-in-the-loop workflow to annotate fingering from the captured video.

This paper introduces an integrated framework aimed at simplifying the dataset acquisition pipeline. By addressing the challenges of data synchronization and fingering annotation, we anticipate our work can contribute to the creation of large-scale multimodal piano datasets and support the empirical research that relies upon them.

\begin{figure}
    \centering
    \includegraphics[width=\linewidth]{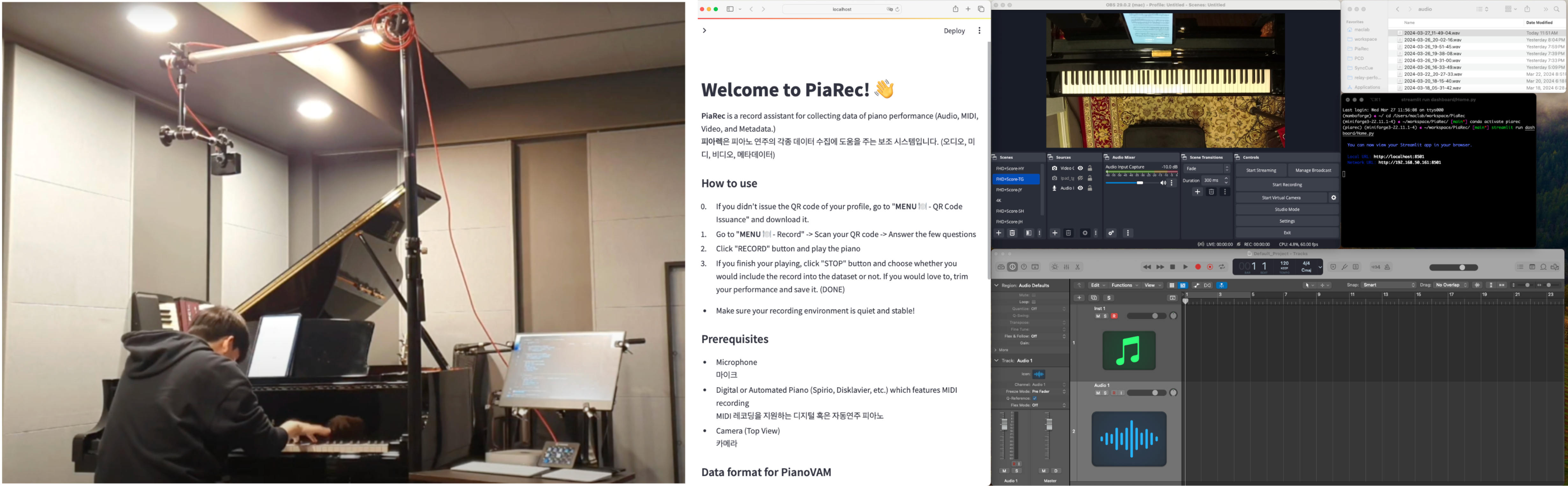}
    \caption{PiaRec system in action, showing (left) the physical recording setup and (right) the PiaRec interface orchestrating OBS Studio and Logic Pro.}
    \vspace{-5mm}
    \label{fig:piarec}
\end{figure}

\section{PiaRec: GUI for Data Acquisition}


PiaRec is a system designed to automate the synchronized acquisition of piano performance data, including audio, video, MIDI, and metadata. It features a Graphical User Interface (GUI) built with Python and Streamlit, which leverages the PyAutoGUI library to directly control external software like Logic Pro and OBS Studio, thereby eliminating manual synchronization errors. Notably, its modular design ensures extensibility, allowing it to be flexibly adapted for use with other Digital Audio Workstations and video capture systems.

\subsection{Workflow and Key Features}
PiaRec is centered around a web dashboard and a QR code-based control system. A first-time user completes a one-time registration on the ``Registration'' tab to generate three QR codes: 
\begin{inparaenum}[(i)]
\item a \emph{Profile} code for user identification, 
\item a \emph{Play} code to initiate recording, and 
\item a \emph{Stop} code to terminate.
\end{inparaenum}

For each recording session, the user inputs performance-specific metadata (e.g., composer, piece title) on the ``Record'' tab. Subsequently, scanning the \emph{Profile} and \emph{Play} codes triggers the automated, simultaneous recording of all data streams in both Logic Pro and OBS Studio. The session concludes upon scanning the \emph{Stop} code, which terminates the session and saves the raw files.

Once the capture is complete, PiaRec performs its automated post-processing. It synchronizes the data streams by cross-correlating the audio from the different sources using the \texttt{numpy.correlate} function to find a precise time offset. This offset is used to trim the MIDI file, aligning it with the audio-visual data. Finally, all metadata is packaged with the synchronized files to create a well-structured data entry. 

\begin{figure}
    \centering
    \includegraphics[width=\linewidth]{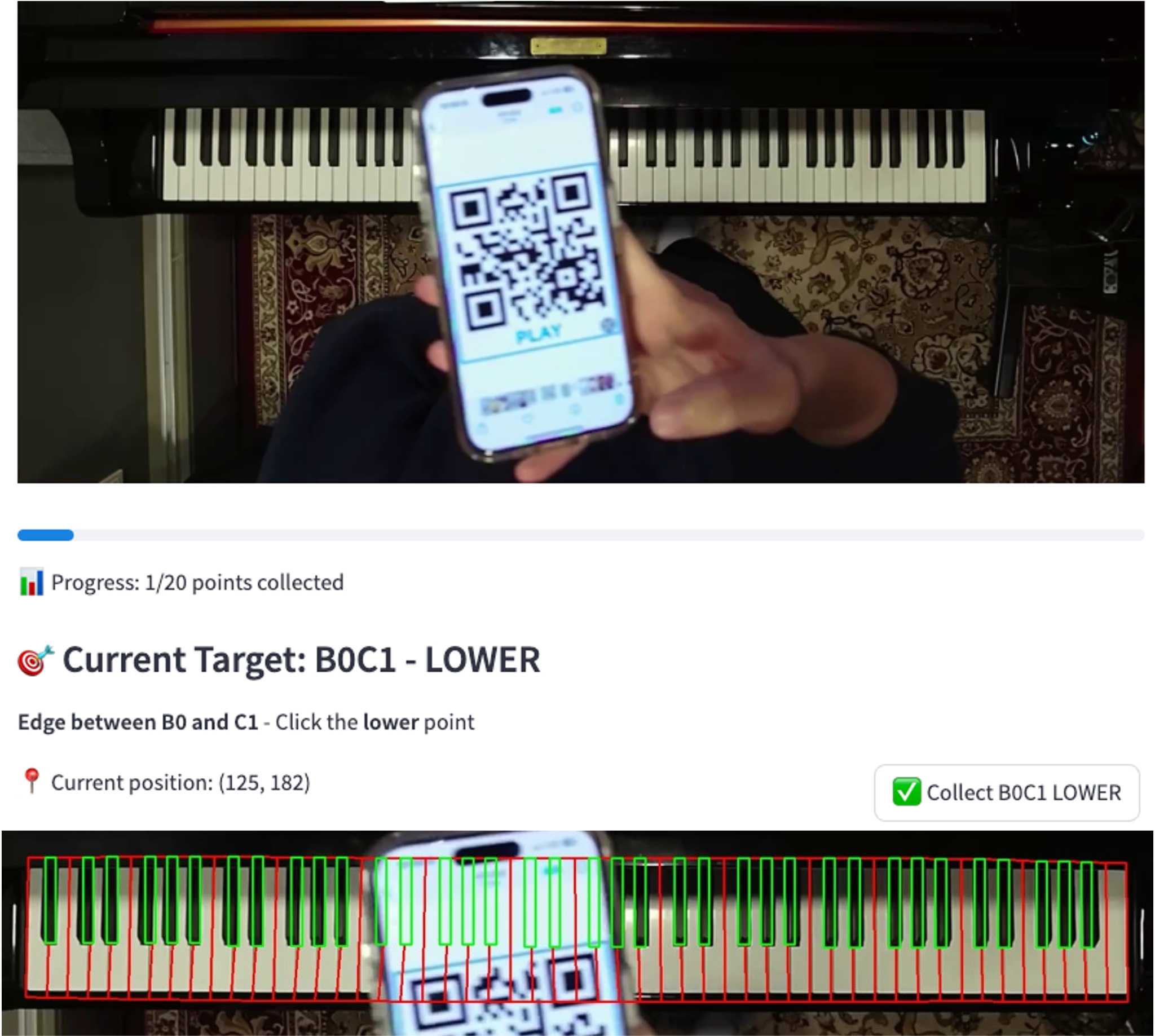}
    \caption{ASDF interface for spatial calibration.}
    \label{fig:asdfkeyboard}  
    \vspace{-5mm}
\end{figure}

\section{ASDF: GUI for Fingering Annotation}
The semi-Automatic System for Detecting Fingering (ASDF) is a toolkit for efficient piano fingering annotation, implemented with the Streamlit framework. It provides an interactive annotation interface for the hybrid workflow proposed by Kim et al.~\cite{kim2025pianovam}, combining an automated fingering detection algorithm with an intuitive interface for human verification.
\subsection{Workflow and Interface Design}
\textbf{Data Preprocessing:} A user begins by loading a performance video and its corresponding MIDI into the system. The first step is the spatial calibration of the keyboard area, performed via the ``Keyboard Detection'' tab. Here, the user defines the specific piano key locations within the video frame. This allows ASDF to map the 88 key regions and apply a heuristic correction for lens distortion by linear interpolation between keystones (see \figref{fig:asdfkeyboard}). Next, the user initiates hand data extraction from the ``Generate Mediapipe Data'' tab. This backend process leverages the Mediapipe Hands \cite{arXiv20Zhang} and the floating hand detection algorithm from Kim et al.~\cite{kim2025pianovam} to generate and save frame-wise skeleton data.\\
\textbf{Automated Candidate Suggestion:} Once pre-processing is complete, the user triggers the automated fingering analysis from the ``Pre-labeling'' tab. With a single action, the GUI executes Kim et al.'s Fingering Candidate Selection Algorithm \cite{kim2025pianovam} described in Algorithm \ref{alg:fingeringalgo}. This algorithm processes the resulting MIDI and hand skeleton data to assign a likelihood score to each finger for every note, generating a set of probable fingering candidates.\\
\textbf{Interactive Annotation and Verification:} The core function of ASDF lies in its main ``Labeling'' tab, which is designed for efficient human-in-the-loop verification. This interface presents a synchronized, multi-panel view containing: 
\begin{inparaenum}[(i)] 
\item the performance video, 
\item a piano roll visualization of the MIDI notes, and 
\item a translucent overlay of the detected hand skeletons on the video. 
\end{inparaenum}
 Notes with a single, high-confidence candidate are pre-labeled, while notes with low confidence or multiple competing candidates are highlighted for manual review. A user can then click on any note in the piano roll to instantly navigate the video to that moment, visually verify the action, and assign or correct the fingering label with a simple input. This design significantly accelerates the annotation process by focusing human effort precisely where it is most needed.

\begin{algorithm}[t]
\caption{Fingering Candidate Selection Algorithm}\label{alg:fingeringalgo}
\begin{algorithmic}
\State $N = \text{Total number of notes}$
\State $K(n) = \text{Keyboard area of $n$th note}$
\State $I(n) = \text{Interval of video frames of $n$th note played}$
\State $H(f,i) = i\text{th finger location info of frame $f$,} $ but fingers of floating hands are not contained
\State $w = \text{width of a key}$ 
\State $S_n =$ Score of each finger likely playing $n$th note
\For{$n < N$}
\State $S_n \gets (0,\cdots,0)\in \mathbb{R}^{10}$
\For{$f \in I(n)$, $i<10$}
\If{\text{each }$H(f,i) \in K(n)$}
\State $S_n\gets S_n + \chi_i$ \Comment{$\chi_i$ = $i$th unit vector}
\ElsIf{$0<d(H(f,i),K(n))_{\mathbb{R}^2}< w$}
\State $S_n \gets S_n + \left(\frac{1-d(H(f,i),K(n))_{\mathbb{R}^2}}{w}\right)^{2}$
\EndIf
\EndFor
\EndFor
\For{$n<N$}
\If{$\exists i \ s.t. \ S_n \cdot \chi_i > 0.5|I(n)|$}
\If{$\exists ! \ i$}
\State Finger $i$ is the only candidate for $n$th note
\ElsIf{$\exists ! \ i \ s.t. \ S_n \cdot \chi_i > 0.8|I(n)|$}
\State Finger $i$ is the only candidate for $n$th note
\Else \text{ }  Multiple candidates for $n$th note
\EndIf
\Else \text{ } No candidate for $n$th note
\EndIf
\EndFor
\end{algorithmic}
\end{algorithm}

\section{Conclusion}
We presented PiaRec and ASDF, the web toolkits designed to lower the significant barriers to creating richly annotated, multimodal piano performance datasets. This integrated pipeline streamlines the entire workflow~---from synchronized data acquisition to efficient fingering annotation---~providing a foundation for data collection that can be expanded in future work.

\section{Acknowledgments}
This research was supported by the National Research Foundation of Korea (NRF) funded by the Korea Government (MSIT) under Grant RS-2023-NR077289 and Grant RS-2024-00358448.

\bibliography{ISMIR2025_LBD}

\end{document}